# Abundances in planetary nebulae: Me 2−1[⋆,⋆⋆]

R. Surendiranath[1], S. R. Pottasch[2], and P. García-Lario[3]

[1] Indian Institute of Astrophysics, Koramangala II Block, Bangalore – 560034, India
[2] Kapteyn Astronomical Institute, PO Box 800, 9700 AV Groningen, The Netherlands
[3] ISO Data Centre. Science Operations and Data Systems Division. Research and Scientific Support Department of ESA. Villafranca del Castillo. Apartado de Correos 50727, 28080 Madrid, Spain

**Abstract.** ISO and IUE spectra of the round planetary nebula Me 2−1 are combined with visual spectra taken from the literature to obtain for the first time a complete extinction-corrected spectrum. With this, the physico-chemical characteristics of the nebula and its central star are determined by various methods including photoionization modeling using Cloudy. The results of the modeling are compared to those derived from a more classical, simple abundance determination approach. A discussion is presented on the validity of the different methods used and assumptions made. Finally, the main results are interpreted in terms of the evolutionary stage of Me 2−1 and its central star.

## 1. Introduction

Me 2−1 is a relatively small, round-shaped, faint planetary nebula, first discovered by Merrill (1942) more than 60 years ago. Observations taken with the Wide Field Planetary Camera 2 (WFPC2) on board HST (see Fig. 1) in several blue and ultraviolet band passes reveal a smooth nebular component extended over ∼9″ with little clumping. The central star, although very faint, is clearly detected at the centre of the nebula in the HST images. Wolff et al. (2000) measured its continuum, deriving a visual magnitude of 18.40 ± 0.05 and a rather high temperature, well above 100 000 K, which we discuss later. Its spectral type is not known as yet.

Me 2−1 (PK 342.1+27.5) is, as the PK number indicates, well above the galactic plane, suggesting that this planetary nebula is located not too far away from us. Later in the paper a distance of 2.3 kpc is suggested. The high galactic latitude can be interpreted as an indication of a low mass progenitor star, which eventually could be reflected in the chemical abundances derived. Statistical distances computed using different methods range between 2 kpc and 7 kpc, which makes an estimation of its actual luminosity and mass uncertain.

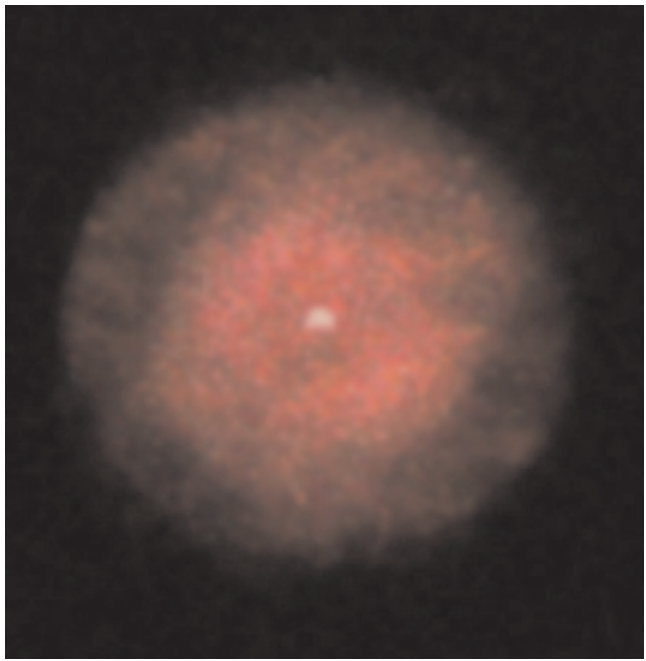

**Fig. 1.** The planetary nebula Me 2−1 as seen with HST. This is a colour-composite image of WFPC2 exposures taken with the *F*547*M* filter (red), the *F*439*W* filter (green) and the sum of the *F*185*W*, *F*218*W*, *F*255*W* and *F*336*W* filters (blue).

The purpose of this paper is to determine the chemical abundances for this nebula and to derive parameters like $T_{\rm eff}$, log $g$, etc., of its central star more accurately than before. This is achieved first by including the ISO (Infrared Space

*Send offprint requests to*: R. Surendiranath,
e-mail: nath@iiap.res.in

⋆ Based on observations with ISO, an ESA project with instruments funded by ESA Member States (especially the PI countries: France, Germany, The Netherlands and the UK) and with the participation of ISAS and NASA. This research has also used archival IUE and HST data.

⋆⋆ Tables 4 and 7 are only available in electronic form at http://www.edpsciences.org

Observatory) spectral data in the analysis. Second, by applying state-of-the-art photoionization modeling, to reproduce the overall spectral energy distribution and the observed nebular emission line intensities from the ultraviolet to the infrared range.

The advantages of incorporating the ISO spectrum in the analysis have previously been discussed (e.g. see Pottasch & Beintema 1999; Pottasch et al. 2000, 2001; Bernard Salas et al. 2001), and can be summarized as follows.

The infrared lines originate from very low energy levels and thus give an abundance which is not sensitive to the temperature in the nebula, nor to possible temperature fluctuations. Furthermore, when a line originating from a high energy level in the same ion is observed, it is possible to determine an effective (electron) temperature $T_e$ at which the lines of that particular ion are formed. When $T_e$ for many ions can be determined, it is possible to make a plot of $T_e$ against ionization potential, which can be used to determine the $T_e$ for ions for which only lines originating from a high energy level are observed. Use of an effective electron temperature takes into account the fact that ions are formed in different regions of the nebula. In this way, possible temperature variations within the nebula can be taken into account.

Use of the ISO spectra have further advantages. One of them is that the number of observed ions used in the abundance analysis is approximately doubled, which removes the need for using large *"ionization correction factors"*, thus substantially lowering the uncertainties in the abundances derived. A further advantage is that the extinction in the infrared is almost negligible, eliminating the need to include large extinction correction factors.

A second method of improving the abundances is by using a nebular model to determine them. This has several advantages. First it provides a physical basis for the electron temperature determination. Secondly it permits abundance determination for elements which are observed in only one, or a limited number of ionic stages. This is true of Mg, S, Cl, K, Ca and Fe which could not be accurately determined without a model. A further advantage of modeling is that it provides physical information on the central star and other properties of the nebula. It thus allows one to take a comprehensive view of the nebula-star complex.

A disadvantage of modeling is that there are possibly more unknowns than observations and some assumptions must be made, for example, concerning the geometry. In our case we will assume that the nebula is spherical and that no clumping exists. The observed round form and smooth emission seen in Fig. 1 make these assumptions reasonable as a first approach. Other assumptions will be discussed in Sect. 5.

This paper is structured as follows. First the spectroscopic data are presented in Sect. 2. In Sect. 3 a preliminary estimate of distance, radius and luminosity of the central star are made. Section 4 discusses a simple method to determine the chemical composition of Me 2−1 and presents the resultant abundances. In Sect. 5 the approach to modeling, its assumptions and results are given. In Sects. 6.1 and 6.2 the nebular density, temperature, and comparison of various determinations of central star temperature are presented. Section 6.3 compares the model and observed spectra. The nebular abundances are presented and discussed in Sect. 7. Section 8 gives a brief sketch of the evolutionary state, and Sect. 9 presents our conclusions.

**Table 1.** Log of IUE NEWSIPS spectra of Me 2−1.

| Data ID | Exp(s) | Disp | Aper | Date |
|---|---|---|---|---|
| SWP05233 | 540 | LOW | LARGE | 1979-05-14 |
| SWP05232 | 5400 | LOW | LARGE | 1979-05-14 |
| LWR04517 | 1800 | LOW | LARGE | 1979-05-14 |
| LWR07737 | 2400 | LOW | LARGE | 1980-05-12 |
| SWP08985 | 3000 | HIGH | LARGE | 1980-05-12 |
| SWP10185 | 6000 | HIGH | LARGE | 1980-09-20 |
| LWR08849 | 1200 | HIGH | LARGE | 1980-09-20 |
| LWR08850 | 7200 | HIGH | LARGE | 1980-09-20 |
| LWP18695 | 2400 | LOW | LARGE | 1990-09-03 |
| SWP39574 | 1200 | LOW | LARGE | 1990-09-03 |

## 2. The spectrum of Me 2−1

In the following we describe the observed ultraviolet, visual and infrared spectroscopic data used in the analysis. A compilation of the extinction-corrected emission line fluxes and identifications are given in Cols. 4 and 8 of Table 7 for a select set of 98 lines.

### 2.1. The IUE ultraviolet spectrum

Ten observations of Me 2−1 were taken with the IUE (International Ultraviolet Explorer), all using the large aperture (10″ × 23″). The log of the IUE observations is given in Table 1. These were retrieved from the STSCI-MAST (IUE) archive, which contains the NEWSIPS (New Spectral Image Processing System) spectra. This processing system incorporated a number of improvements and enhancements in the reduction algorithms and calibrations (see http://archive.stsci.edu/iue/newsips/newsips.html for more details). The available spectra are a combination of low and high resolution observations covering from 1150 Å to 3200 Å and were acquired between 1979 and 1990. Of these, the four high resolution spectra are found to be of poor quality.

We have merged the remaining low resolution (∼6 Å) spectra and extracted the fluxes using IUETOOLS under IRAF. The merging and exposure-weighted averaging yielded a better S/N spectrum. We were able to detect some lines unreported so far; the new lines and their probable identifications are 1574 Å ([Ne V] 1575), 1599 Å ([Ne IV] 1602), 1866 Å (Al III 1863), 2301 Å (C III 2297) and 2784 Å ([Mg V] 2785).

To minimize the errors in the measured fluxes due to saturation effects, the line fluxes of 1547 Å (C IV 1548, 50), 1639 Å (He II 1640) and 1906 Å (C III] 1907, 09) were measured on the shortest exposed spectrum (SWP 05233). A check with Feibelman's (1994) extraction for these strong lines showed good agreement with those from the abovementioned averaged spectrum. Figures 2 and 3 show the final spectra used in this work and Table 7 gives the extracted fluxes.

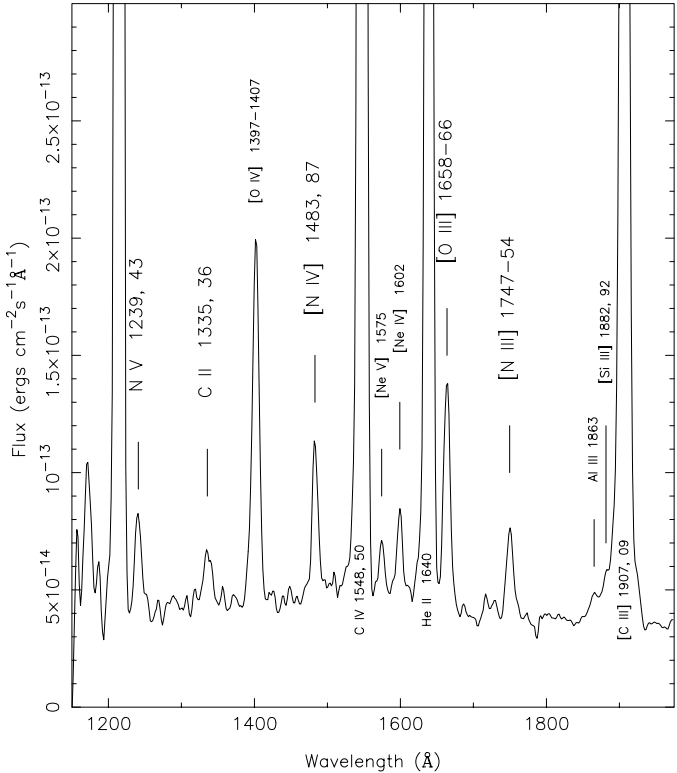

**Fig. 2.** The averaged IUE SWP spectrum of Me 2–1.

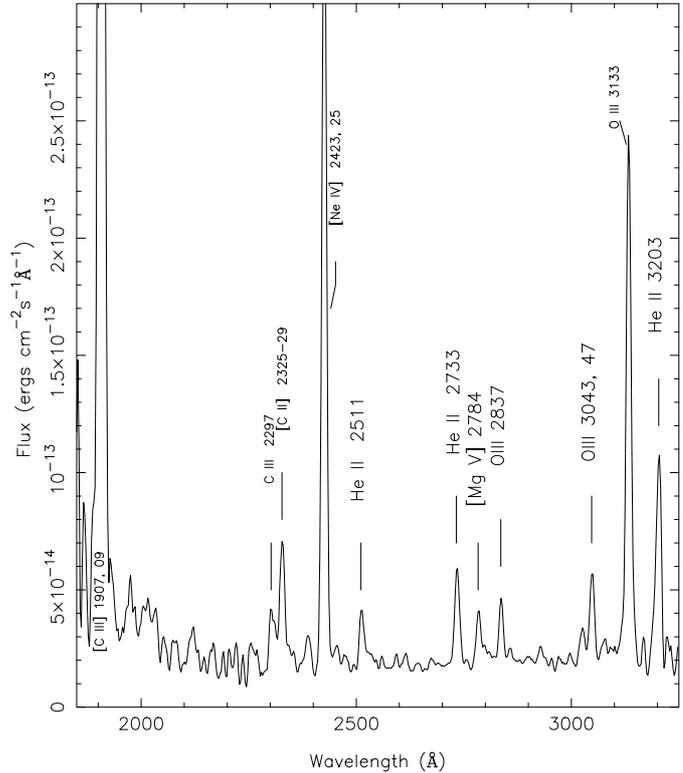

**Fig. 3.** The averaged IUE LWP/LWR spectrum of Me 2–1.

### 2.2. The visual spectrum

Me 2–1 has been optically observed on many occasions in the last two decades (see Aller et al. 1981; Cuisinier et al. 1996; Kingsburgh & Barlow 1994; Moreno et al. 1994). Due to the better consistency in measured fluxes, line identifications and coverage of wavelength range, we decided to use the spectral observations of Aller et al. (1981) and Moreno et al. (1994) for our work, after averaging them. The averaged fluxes are listed in Table 7.

### 2.3. ISO observations

The ISO SWS observations were made on August 5, 1997 (TDT 62803316) with the SWS 02 observing mode (see the ISO Handbook Vol. V – Leech et al. 2003), which provides a spectral resolution of $\lambda / \Delta\lambda \sim 1000$–$2000$. Several small wavelength intervals were observed covering the spectral range from 2 to 37 $\mu$m leading to the detection of 7 nebular emission lines plus Br$_\alpha$ at 4.05 $\mu$m. A set of seven more emission lines that was expected within the observed wavelength range was not detected. We estimated upper limits for them and included them in the analysis. These are marked by the symbol "<" in Table 7.

All the observed lines are unresolved at this spectral resolution. The aperture size of the instrument at these wavelengths ranged from 14″ × 20″ to 20″ × 33″, admitting radiation from the entire nebula. Data reduction of the pipeline products directly retrieved from the ISO Data Archive was carried out using ISAP (ISO Spectral Analysis Package) version 2.1, developed at IPAC (Sturm et al. 1998). First, data points affected by cosmic rays or deviating significantly from the majority were removed through a $\sigma$ clipping. Then, since the spectroscopic measurements consist of several up-down scans done within a given band, these were compared and where no significant differences were found, the measurements in the two directions were averaged. Figure 4 shows the reduced ISO spectra. Since most of the lines are well represented by Gaussian profiles we fitted the line profiles to a Gaussian function and obtained the wavelength of the line centre, the FWHM (full width at half maximum) and the integrated flux of each line using the available standard routines in ISAP.

The uncertainties in the absolute flux calibration are expected to be approximately 20% for the stronger emission lines and about 30% for the fainter ones. The reliability of detection of even the faintest lines has been ascertained by the consistency of the associated Doppler shifts among all lines.

### 2.4. Extinction

There are several methods for estimating the extinction towards planetary nebulae; for example, the comparison of the observed and the theoretical Balmer decrement, the comparison of the radio emission with the H$_\beta$ flux, etc. Aller et al. (1981) derived a value of 0.36 for the extinction constant $c$; Moreno et al. (1994) obtained a similar value of 0.34, both from the Balmer decrement. However, other values can be found in the published literature. Preite-Martinez & Pottasch (1983) give a value of $E_{B-V} = 0.18$ i.e. $c = 0.26$. They obtained this by averaging three values derived i) by comparing the radio flux with the H$_\beta$ flux; ii) from the absorption dip at 2200 Å, and iii) from the Balmer decrement. On the other hand, if we use the $H_\alpha$ to $H_\beta$ ratio from Kaler's (1983) observations we get $c = 0.16$ while using average value from Aller et al. (1981) and Moreno et al. (1994) for the same ratio yields $c = 0.4$.

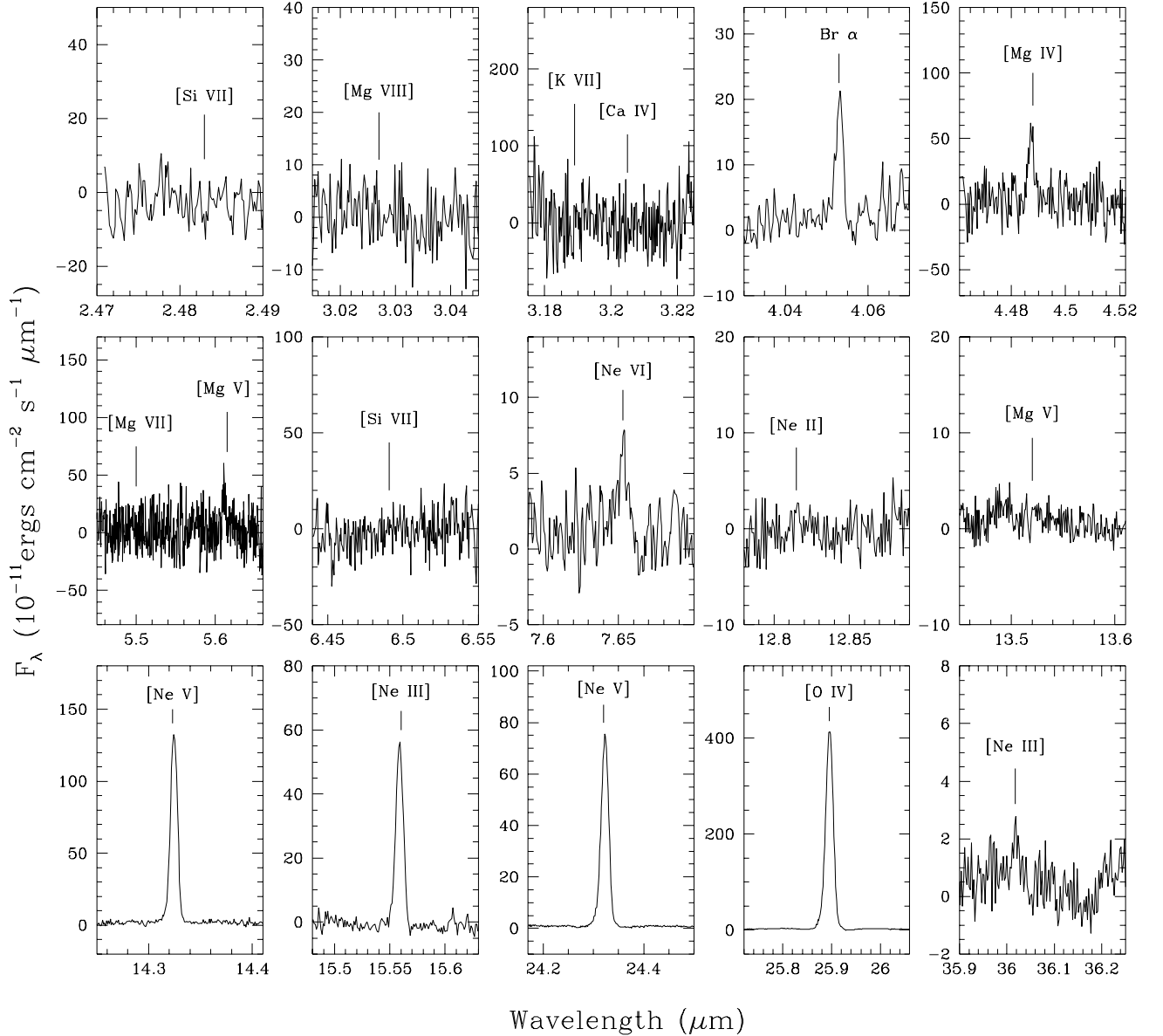

**Fig. 4.** The ISO SWS 02 spectra of Me 2−1.

In view of the wide variations in this value found in the literature, we redetermined the extinction constant. First, we adopted the value of $\log H_\beta = -11.31$ from Kaler (1983), since his observation covers the full diameter of the nebula. Then, we decided that the best way of finding a reliable value for $c$ was to choose a value that would return the unreddened ratios close to the theoretical ones at least for the strong (and therefore well detected) lines. This yielded $c = 0.28$ (i.e. $E_{B-V} = 0.19$), which gives the following theoretical values (unreddened observational values in parenthesis): $H_\alpha/H_\beta = 2.81$ (3.0); $H_\gamma/H_\delta = 1.81$ (1.75); $He^+$ 5412/4686 = 0.077 (0.072) and 4686/1640 = 0.145 (0.150). The extinction derived from HST measurements by Wolff et al. (2000) is $E_{B-V} = 0.15$ which returned the following corrected values for the above-mentioned line ratios in the same order: 3.21, 1.76, 0.074 and 0.174. Extinction is best determined by two points in the spectrum that are well separated in wavelength, preferably one in the UV (i.e. short wavelength) where extinction is maximum and the other at a much longer wavelength where it is much lower. The shortest wavelength included in the HST observations by Wolff et al. (2000) was ∼1900 Å, and the longest one was at ∼5500 Å. All the observations long-ward of 3400 Å could not distinguish among different values of extinction for the various temperatures considered (see Fig. 5, ibid). We thus gave due importance to the line ratio of (4686/1640).

Our new estimate is in good agreement with the one derived by Preite-Martinez & Pottasch (1983), and in the remainder of this paper we will use this value $c = 0.28$ or $E_{B-V} = 0.19$, together with the extinction curves of Seaton (1979) and Fluks et al. (1994). Finally, since Me 2–1 is a high excitation object i.e., $F$(He II 4686 Å) is more than 75% of $F$(H$_\beta$), we corrected the unreddened H$_\beta$ flux for contamination by He$^+$ Pickering $8 \to 4$ transition. All unreddened line fluxes were normalized to this unreddened and uncontaminated H$_\beta$ flux as 100 units; these are given in Table 7.

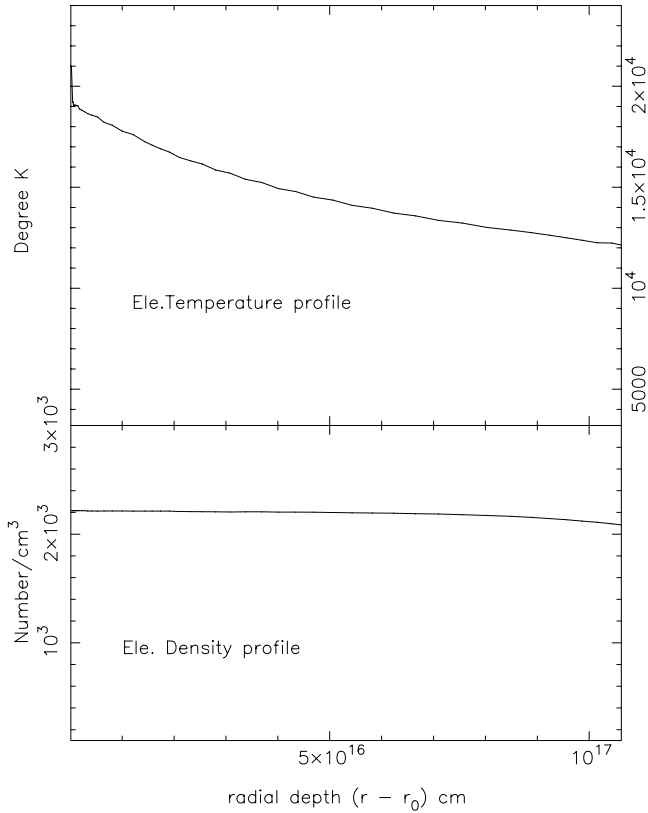

**Fig. 5.** $T_e$ and $N_e$ across the nebula.

## 3. Radius and luminosity of the central star

These quantities are strongly dependent on the distance of the nebulae which is difficult to obtain accurately. By equating the ⟨rms⟩ density with the forbidden line density (see Sect. 4.1) a value of $d = 2.3$ kpc is found and this will be the value used when necessary throughout the rest of this paper. That the distance is at the low end of the statistical distances, taken together with the high galactic latitude, may indicate that the nebula is formed from a nearby low mass star. This value, however, has an uncertainty which could be larger than 40%. This leads to stellar radii $R/R_\odot = 0.028$. Using a value for temperature $T = 140\,000$ K (which is the "Stoy" temperature $T_{\rm EB}$; see Sect. 6.2.1), we obtain the stellar luminosity $L/L_\odot = 260$.

It is also possible to obtain the stellar luminosity from the nebular H$_\beta$ luminosity, since there is a direct relationship between the number of ionizing photons and the number of H$_\beta$ photons in the case in which the nebula absorbs all the ionizing photons emitted by the star. A mathematical formulation of this can be found in Pottasch & Acker (1989). It yields the following luminosity: $L/L_\odot = 240$. This is roughly the same value found above and indicates that most of the ionizing photons must actually be absorbed in the nebula. However, it does not give any information about the distance, since both formulations have the same distance dependence.

**Table 2.** Electron density indicators in Me 2–1.

| Ion | Ioniz. Pot.(eV) | Lines Used | Observed Ratio | $N_e$ (cm$^{-3}$) |
|---|---|---|---|---|
| [S II] | 10.4 | 6731/6716 | 1.37 | 1900 |
| [O II] | 13.6 | 3626/3729 | 1.24 | 1800 |
| [Cl III] | 23.8 | 5538/5518 | 0.93 | 1600 |
| [C III] | 24.4 | 1907/1909 | 1.27 | 1200 |

## 4. Chemical composition of Me 2–1 from the simplified analysis

The method of analysis is the same as used in the papers cited in the introduction. First the electron density and temperature as functions of the ionization potential are determined. Then the ionic abundances are determined, using density and temperature appropriate for the ion under consideration, together with Eq. (1). Then the element abundances are found for those elements for which a sufficient number of ionic abundances have been derived.

### 4.1. Electron density

The ions used to determine $N_e$ are listed in the first column of Table 2. The ionization potential required to reach that ionization stage, and the wavelengths of the lines used, are given in Cols. 2 and 3 of the table. Note that the wavelength units are Å when 4 ciphers are given and microns when 3 ciphers are shown. The observed ratio of the lines is given in the fourth column; the corresponding $N_e$ is given in the fifth column. The temperature used is discussed in the following section, but is not important since these line ratios are essentially determined by the density.

The electron density appears to be about 1700 cm$^{-3}$. There is no indication that the electron density varies with ionization potential in a systematic way. It is interesting to compare this value of the density with the ⟨rms⟩ density found from the H$_\beta$ line. This depends on the distance of the nebula which is not accurately known, and on the angular size of the nebula. Because of the distance uncertainty, we shall turn the calculation around, and compute what the distance will be for an ⟨rms⟩ density of 1700 cm$^{-3}$ in a sphere of radius 4.5″, that emits the H$_\beta$ flux given above. This yields a distance of 2.3 kpc. This value will be used in further computations in this paper.

### 4.2. Electron temperature

A number of ions have lines originating from energy levels far enough apart that their ratio is sensitive to the electron temperature. These are listed in Table 3, which is arranged similarly to the previous table. The electron temperature is found to increase as a function of ionization potential. There is some scat-ter. The [Ne V] temperature is high which might indicate that the intensity of the line at $\lambda$ 3425 Å has been overestimated.

**Table 3.** Electron temperature indicators in Me 2−1.

| Ion | Ioniz. Pot.(eV) | Lines Used | Observed Ratio | $T_e$ (K) |
|---|---|---|---|---|
| [N II] | 14.5 | 5755/6584 | 0.0223 | 11 500 |
| [S III] | 23.3 | 6312/9531 | 0.0494 | 12 800 |
| [Ar III] | 27.6 | 5192/7136 | 0.0120 | 12 000 |
| [O III] | 35.1 | 4363/5007 | 0.0155 | 13 500 |
| [O III] | 35.1 | 1663/5007 | 0.032 | 13 000 |
| [Ne III] | 41.0 | 3868/15.5 | 1.375 | 12 700 |
| [O IV] | 54.9 | 1400/25.9 | 0.0943 | 16 000 |
| [Ne IV] | 63.4 | 2424/4725 | 0.0109 | 18 000 |
| [Ne V] | 97.1 | 3425/24.3 | 1.13 | 21 000 |
| [Mg V] | 109.3 | 2784/5.61 | 0.585 | 17 000 |

### 4.3. Ionic and element abundances

The ionic abundances have been determined using the following equation:

$$\frac{N_{ion}}{N_p} = \frac{I_{ion}}{I_{H_\beta}} N_e \frac{\lambda_{ul}}{\lambda_{H_\beta}} \frac{\alpha_{H_\beta}}{A_{ul}} \left(\frac{N_u}{N_{ion}}\right)^{-1} \quad (1)$$

where $I_{ion}/I_{H_\beta}$ is the measured intensity of the ionic line compared to $H_\beta$, $N_p$ is the density of ionized hydrogen, $\lambda_{ul}$ is the wavelength of this line, $\lambda_{H_\beta}$ is the wavelength of $H_\beta$, $\alpha_{H_\beta}$ is the effective recombination coefficient for $H_\beta$, $A_{ul}$ is the Einstein spontaneous transition rate for the line, and $N_u/N_{ion}$ is the ratio of the population of the level from which the line originates to the total population of the ion. This ratio has been determined using a five level atom.

The results are given in Table 4, where the first column lists the ion concerned, and the second column the line used for the abundance determination. The third column gives the intensity of the line used relative to $H_\beta = 100$. The fourth column gives the electron temperature used, which is a function of the ionization potential and is taken from Table 3. The ionic abundances, are in the fifth column, while the sixth column gives the Ionization Correction Factor (ICF). This has been determined empirically. Notice that the ICF is unity for helium, carbon, nitrogen, oxygen and neon because all important stages of ionization have been observed. The ICF for the other elements has been determined by comparing the observed ionization stages as a function of ionization potential with those elements where all important ionization stages are present, especially nitrogen and neon. The three ionization stages in both argon and potassium are the most important and thus justify the use of an ICF close to unity. For sulfur the importance of the missing ionization stages is difficult to judge, therefore an empirical ICF is very uncertain. This is less true of magnesium and chlorine, which might still be uncertain by a factor of two. Only one stage of ionization has been observed in silicon and calcium for which only a model approach can give a trustworthy solution. Iron was not attempted with this approach. The element abundances are given in the last column of the table.

The helium abundance has been derived using the theoretical work of Benjamin et al. (1999). For recombination of singly ionized helium, most weight is given to the $\lambda$ 5875 Å line, because the theoretical determination of this line is the most reliable.

The abundances in Me 2−1 are in general very similar to solar abundances. The oxygen abundance is almost solar. Neon, argon and chlorine are slightly lower. Nitrogen is even lower than in the Sun, indicating that very little dredge-up has taken place.

#### 4.3.1. Recombination line abundances

The only recombination line in the spectrum is the C II line at $\lambda$ 4267 Å. Using the observed ratio of this line to $H_\beta$ of $4.1 \times 10^{-3}$ and the effective recombination coefficient to form this line given by Davey et al. (2000) at an electron temperature of $T = 12 500$, we obtain a value of $C^{++} = 4.7 \times 10^{-4}$. This is very close to the value of $4.98 \times 10^{-4}$ obtained from the collisionally excited line at $\lambda$ 1909 Å. The two values are equal within the uncertainties of the observations.

## 5. Model

There is ample motivation to take a broader view of the nebula-star complex. As we noted at the outset, the abundances of certain elements could not be derived from the simple approach made above. As regards the central star, though Wolff et al. (2000) measured its continua in various filter bands from the ultraviolet to optical region, there is some uncertainty about the $T_{eff}$. Wolff et al. (2000) quote a range from 119 000 K to 230 000 K. Although they have fitted a black body curve over the HST observations to get a temperature, they acknowledge the difficulty in discriminating between different temperatures and concede that their fits are a more sensitive determination of reddening. The reddening value derived by them was lower ($E_{B-V} = 0.15$) than our determination. We have shown that this value could not be used.

Various methods exist to determine the temperature of the central stars of planetary nebulae. The most important of these are the 1) Zanstra method; 2) Energy balance method; 3) NLTE study of the central star spectrum; 4) modeling the degree of nebular ionization. The first three methods are known to be dependent on various uncertain assumptions. They therefore give conflicting and controversial temperatures. The fourth method, for which the ISO observations are needed, is a comprehensive procedure and can give a measure of self-consistency in the results. In the optical and in the UV only "low" ionization stages are seen, which do not provide enough information to distinguish temperatures above 100 000 K. However, with ISO SWS we can cover lines of much higher ionization potential, up to 303 eV.

Modeling the nebula-star complex will allow us to characterize the central star's atmosphere (i.e., $T_{eff}$, $\log g$ and luminosity), to determine the distance and other nebular properties like its composition. This method can take into account the presence of dust and molecules in the nebular material and thus is very comprehensive in approach. While the line ratio method is simple and fast, the ICFs rest on uncertain physics and often one needs to consider all details of observations and theory since every parameter is inter-related and dictated by nebular astrophysics and astrochemistry, both locally and globally. To this end, modeling is effective and the whole set of parameters is determined in an unified way, assuring self consistency. In this way one gets a good physical insight into the PN, the method and the observations.

**Table 5.** Parameters representing the best-fit model.

| Parameter | Value |
|---|---|
| *CSPN* | |
| *Model atmosphere* | |
| $T_{\rm eff}$ | 170 000 K |
| log $g$ | 7.3 |
| Luminosity | 490 $L_\odot$ |
| | |
| *Nebula* | |
| Const. density | $N_H = 1.840 \times 10^3$ cm$^{-3}$ |
| Abundance | H    He    C    N |
| | 12.000  11.000  8.8325  7.8261 |
| | O    Ne    Mg    Si |
| | 8.7076  7.9031  7.3802  6.7782 |
| | S    Cl    Ar    K |
| | 6.5441  5.2553  6.2041  4.6990 |
| | Ca    Fe |
| | 6.3600  5.7500 |
| Size | 8″.7 (diameter) |
| Distance | 2.300 kpc |
| Inner radius | 4.365158e16 cm (∼0.0141 pc) |
| Outer radius | 1.496725e17 cm (∼0.0485 pc) |
| Filling factor | 1 |

It is with this in mind that we constructed a photoionization model for Me 2−1 with the code Cloudy, version 96.04 i.e., 96 beta 5 (Ferland 2001).

### 5.1. Assumptions in the model

Though Cloudy allows incorporation of dust grains and a network of molecules and ions, we have not included either. The observed IRAS fluxes are rather low and no molecular emission has been reported. Kastner et al. (1996) did not detect 2.122 µm emission of molecular hydrogen in Me 2−1, and Payne et al. (1988) report non-detection of OH. We considered only gas phase abundances of the various elements seen in the spectra (Table 5).

We examined the HST images of Me 2−1(see Fig. 1) available in the HST Data Archive and found that the image has a sharp-edged round morphology. We determined an angular diameter of 8″.7 from these images, and used this value for modeling. Several earlier determinations from radio observations gave a smaller diameter (∼7″). The smooth distribution of light and absence of marked clumping in the images indicated that a constant density model would be sufficient, though Cloudy permits a variety of density distributions. We considered a spherically symmetric and static nebula. All the properties derived for Me 2−1 in Sects. 1−4 were used as initial values in the modeling. For elements whose abundances were not known a priori, typical PN abundances were inserted. To represent the central star's energy distribution in the model, we chose the Rauch (2003) model atmospheres. These NLTE model atmospheres include metal-line blanketing, i.e., include light metals as well as the iron-group (Sc, Ti, V, Cr, Mn, Fe, Co and Ni) opacities with millions of lines following Kurucz's line lists. They are computed for both solar and halo abundances of the central star. Therefore they are ideal for our purpose.

### 5.2. Results

The general method of application of the code Cloudy to model Me 2−1 is the same as in Surendiranath (2002); we ran a number of models and each time the output was carefully scrutinized before running the next model with changed input parameters. We tried to match the observed spectral fluxes of about a hundred lines. We relied on physical intuition rather than any of the optimization techniques provided in Cloudy. Cloudy, in its model output, gives the incident and transmitted energy of the central star (CSPN) radiation as well as many other quantities including nebular spectrum. Initially we attempted to fit not only the observed nebular emission spectrum but also the CSPN continua as measured by Wolff et al. (2000). We tried with various model atmospheres from Rauch (2003) and kept the luminosity of the CSPN at the value given by Zhang & Kwok (1991). They actually give the total integrated flux normalized to H$_\beta$. Taking the value of the H$_\beta$ flux used in our work, we obtain a value of ∼240 $L_\odot$. Our earlier estimated value (Sect. 3) is also close to this.

After a number of numerical experiments we found that it was not possible to get at the same time a good match of the model nebular emission spectrum with the observations and of the transmitted continua with the HST observations. This effectively meant that we needed a very hot CSPN. When CSPNs of lesser $T_{\rm eff}$ were tried we could match the transmitted continua with the HST observations only for certain models but could not get a matched nebular spectrum at the same time. Thus we did the modeling without imposing a good match between the model continua and the HST observations as a necessary condition. In this way we could obtain a well matched nebular emission spectrum. Further discussion on the stellar continua is given later. The model value of the absolute $V$ magnitude is 6.43 which (for the distance and extinction used as input) yields $m_v$ of 18.85. This differs from the value (18.40) given by Wolff et al. (2000) as measured with HST.

Based on the sensitivity of the model results to changes in the input parameters, we estimate the accuracies in the parameters (quoted in Table 5) as follows: ($N_{\rm H}$, ±30 cm$^{-3}$; luminosity, ±30 $L_\odot$; $T_{\rm eff}$, ±10 000 K; distance, ±0.1 kpc; abundances, ±30%).

Table 5 gives the input parameters of the best matched model, and the corresponding output spectral fluxes are compared to the observed ones in Table 7. The abundances in Table 5 are given on a logarithmic scale of log ($N_{\rm H}$) = 12.0.

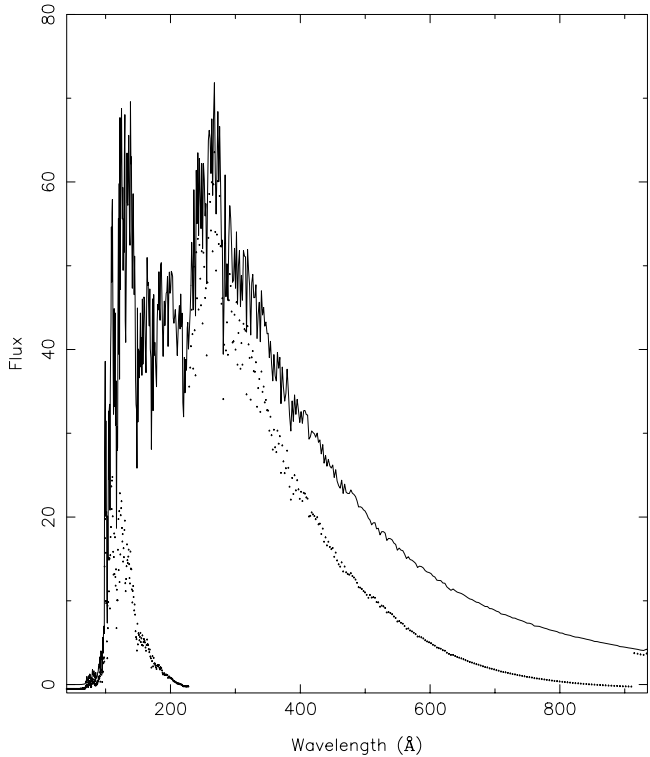

**Fig. 6.** Stellar ionizing radiation – Incident (solid line) and transmitted (dotted line); the $y$-axis values are suitably scaled.

For verification by users of Cloudy, we give the parameters to a higher number of significant figures (exactly as we used) than would be dictated by the accuracy involved in the method.

## 6. Discussion

The best-fit model input parameters are our final values for the characteristics of the PN and its central star.

### 6.1. Nebular density and temperature

The mean nebular electron density from the model is $\sim 2100$ cm$^{-3}$ which is in broad agreement with values derived from [O II] and [S II] in Table 2. The density from [Cl III] and [C III] are uncertain. The run of electron density and temperature across the nebula is shown in Fig. 5. It can be seen that the density at the outer edge of the nebula is very high indicating a high level of ionization. This means that the stellar photon absorption by the nebular material is not total. Indeed we found that the nebula is optically very thin to ionizing radiation (see Fig. 6). We could not get an energy conserving model for the PN such that the electron density at the outer edge would be very low i.e., energy conserving models did not produce spectra that match observations. Observations in the wavelength range of 70 to 700 Å, like those from the former EUV Explorer mission of NASA, would be ideal to settle this issue. The run of electron temperature across the nebula in Fig. 5 is in good agreement with the values obtained from line ratios (Table 3). The nebular radio emission at 5 GHz from the model is $\sim 37$ mJy which compares well with the observed value of 38 mJy quoted by Pottasch (1983). The ionization structure of various elements is shown in Figs. 7–9.

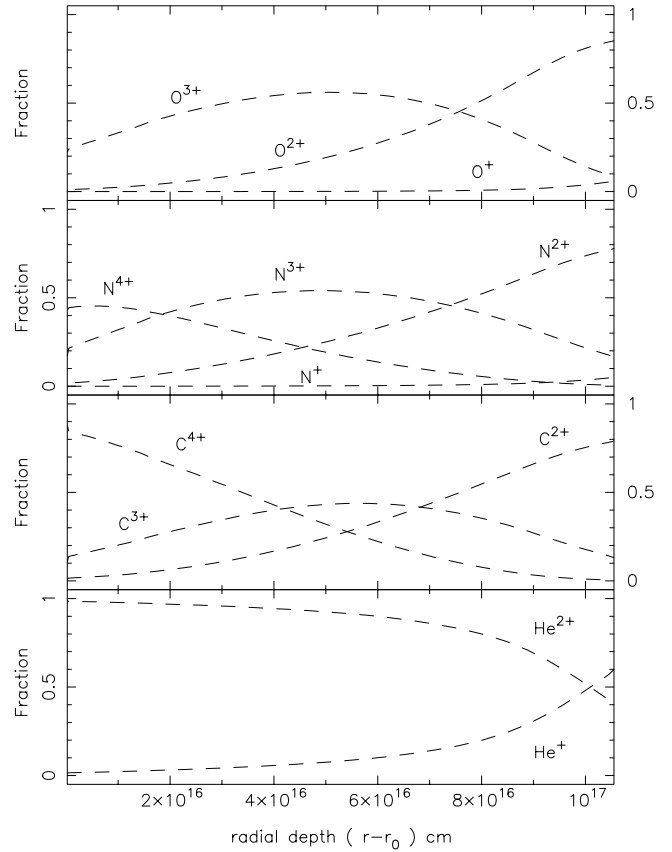

**Fig. 7.** Ionization structure of He, C, N and O. ($r$ = outer radius; $r_0$ = inner radius.)

**Table 6.** Central star continua from model and observation.

| Wavelength (Å) | Model $\nu f_\nu$ | Observation $\nu f_\nu$ |
|---|---|---|
| 1900 | 2.5e-11 | 3.4e-11 |
| 2117 | 1.8e-11 | 3.1e-11 |
| 2545 | 1.1e-11 | 1.5e-11 |
| 3317 | 4.8e-12 | 6.6e-12 |
| 4283 | 2.2e-12 | 3.2e-12 |
| 5446 | 1.1e-12 | 1.5e-12 |

The continua are in units of erg cm$^{-2}$ s$^{-1}$.

### 6.2. Central star

The central star is quite hot in our model with its $T_{\rm eff}$ at 170 000 K. We have experimented with abundances of both solar and halo composition; the two do not produce very different output spectra but the model presented here is the one with solar composition since the fit was better for some lines. If we assume that the mass is 0.6 $M_\odot$, then for the model log $g$ of 7.3, we get $R_\star = 0.029\,R_\odot$. Table 6 compares the continuum flux from the model with the HST observations.

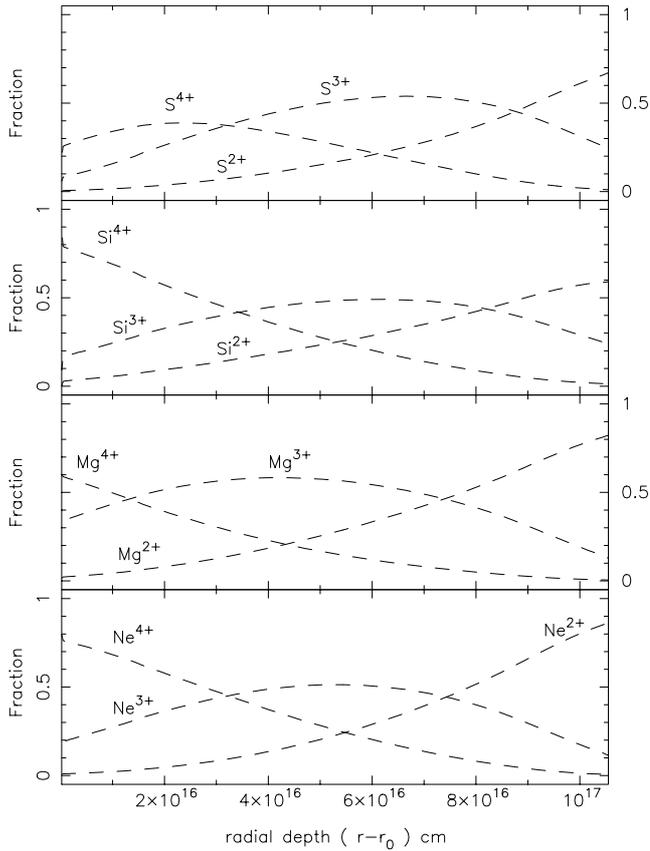

**Fig. 8.** Ionization structure of Ne, Mg, Si and S. ($r$ = outer radius; $r_0$ = inner radius.)

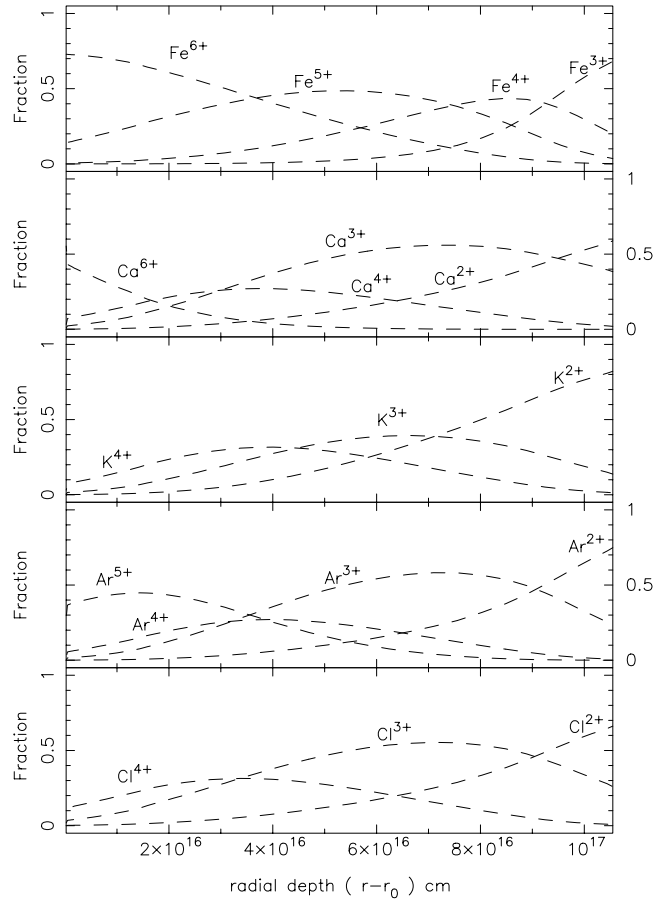

**Fig. 9.** Ionization structure of Cl, Ar, K, Ca and Fe. ($r$ = outer radius; $r_0$ = inner radius.)

### 6.2.1. Comparison of central star temperature determinations

As mentioned in Sect. 1, this nebula is excited by a faint star of unknown spectral type. Using the visual magnitude of 18.40 (obtained from the HST measurements of Wolff et al. 2000) and the H$_\beta$ flux given above, the hydrogen Zanstra temperature $T_z$(H) is about 130 000 K. The doubly ionized helium Zanstra temperature $T_z$(HeII) is about 145 000 K. The "Stoy" or Energy Balance temperature can also be found from the above data. The value of the ratio of "forbidden line emission" (including all collisionally excited emission) to H$_\beta$ is about 58, which leads to an energy balance temperature $T_{EB}$ of 142 000 K, assuming blackbody emission and Case II (the nebula is optically thin for radiation which will ionize hydrogen, but optically thick for radiation short-ward of the ionized helium limit, see Preite-Martinez & Pottasch 1983). If Case III is used (also optically thick for radiation that will ionize hydrogen) we find $T_{EB}$ = 130 000. The best photoionization model gives $T_{eff}$ = 170 000 K. Lower temperatures did not result in a good fit to the observed spectrum.

### 6.3. Comparison of model spectrum with observation

As mentioned earlier we have attempted to obtain a good match for about 100 observed lines shown in Table 7. Cloudy computes by default the fluxes of continuum at various wavelengths and a very large number of emission lines (nearly 2000) in its output spectrum (a copy can be obtained from the first author). The notation for line identification is by a label as per Cloudy. This makes identifying any line in Cloudy's huge line list ($\sim 10^6$) easier (see notes at the bottom of Table 7). The match between model and observation is in general very good. There are some deviant lines. H$_\alpha$ is weaker in the model while some other Balmer lines are stronger than the observations. We could not get any closer match than this. The observed flux of H$_\alpha$ seems to us abnormally high. The line 3355 Å is actually a composite of He I, [Cl III] and [Fe III]. The model flux for the second line is 0.0718. The last line is not included in Cloudy, so the disparity is understandable. The N V line at 1240 Å is deficient in the model. This is so since a fraction of N is pumped up to N VI as there is an ample supply of high energy stellar photons. This could not be the case of 1402 of [O IV], since the far-infrared line at 25.88 $\mu$m of the same ion matches reasonably with the observed value to within 15% accuracy. Neon lines in general behaved rather erratically in all the models we tested and we do not understand their deviant behaviour. There are a few observed weak lines of potassium (4163 Å and 6102 Å), found both in Aller et al. (1981) and Moreno et al. (1994), but these lines are not included in Cloudy's default list. Cloudy has procedures to introduce new lines given by users, but for this work, we have used only the default list throughout. Concerning those 7 lines which have not been detected

**Table 8.** Abundances in Me 2−1.

| Element | Abundance | | | |
|---|---|---|---|---|
| | ICF | Model | Moreno | AKC |
| He | 0.1 | 0.1 | 0.1 | 0.102 |
| C(–4) | 7.0 | 6.80 | | 6.4 |
| N(–5) | 5.1 | 6.70 | 16. | 9.1-16. |
| O(–4) | 5.3 | 5.10 | 7.4 | 5.3 |
| Ne(–5) | 9.3 | 8.00 | 9.8 | 15. |
| Mg(–5) | 2.0 | 2.40 | | |
| Si(–6) | 10. | 6.00 | | |
| S(–6) | 9.1 | 3.50 | 2.9 | 16. |
| Cl(–7) | 1.8 | 1.80 | 3.1 | 1.8 |
| Ar(–6) | 1.6 | 1.60 | 1.7 | 2.5 |
| K(–8) | 5.0 | 5.00 | | 26. |
| Ca(–6) | | 2.29 | | 0.1 |
| Fe(–7) | | 5.62 | | |

The abundances listed as "Moreno" are taken from Moreno et al. (1994), those listed as "AKC" are taken from Aller et al. (1981).

in ISO spectra but for which we have given the upper limits, the model fluxes have been either zero or much lower than the limits, except in one case, where it slightly exceeded the limit. These lines are indicated by the symbol "<" in Table 7.

## 7. Nebular abundances

Barring potassium, all elements have been represented by at least one or more lines in the modeling process. The abundances derived by the two methods are compared in Table 8. The abundances found in the model and those found in the simpler approach (ICF in the table) are the same within the errors present in the two analyzes, in the case of helium, carbon, nitrogen, oxygen, neon, argon and chlorine. For the other elements, (Mg, Si, S, K, Ca, and Fe) the ionization correction factors are too uncertain for the simpler approach to give accurate results, and we recommend using the abundances derived from the model.

Two earlier abundance determinations are also listed in Table 8. Those of Moreno et al. (1994) are based only on the visual spectrum so that no carbon abundance can be determined. Only one nitrogen line could be observed ([N II]) so that an ionization correction factor of more than two orders of magnitude was required, making their nitrogen abundance very uncertain. Aller et al. (1981) included the IUE ultraviolet spectrum in their analysis. Their nitrogen abundance is higher because they used an electron temperature determined from the [O III] lines for the higher nitrogen stages of ionization. Both our simplified analysis and our model analysis show that a higher electron temperature should have been used.

## 8. Evolutionary state

The oxygen abundance is similar to Solar and it is therefore likely that the original composition of the star was nearly Solar. The nitrogen abundance is now also similar to Solar, therefore probably not very much nitrogen was formed in the course of stellar evolution by the so called "second dredge-up". This means that the initial mass of the star was lower than 2.4 solar masses. The large carbon abundance was probably produced during the "third dredge-up", which also increased helium by a small amount. Both the "first" and the "third dredge-up" do not cause a substantial increase in nitrogen abundance. This would be compatible with an initial stellar mass of about 1.5 solar masses. This would also be compatible with the position of the nebula about 1 kpc above the galactic plane.

## 9. Conclusion

A complete spectrum from the UV to far-IR of Me 2−1 has been presented by combining ground-based optical spectra taken from the literature and space-based spectra recently obtained by ISO and IUE. This information, complemented with the morphology and size observed in the available HST images, has been used to get a comprehensive view of the nebula-star complex by means of a detailed photoionization modeling. The set of parameters derived by this method, as given in Table 5, is the best self-consistent set for both the CSPN and the nebula as explained earlier. Me 2−1 seems to have originated from a low mass star (only slightly more massive than the Sun) that initially had a nearly Solar composition, and has undergone the "first" and the "third dredge-up" events.

*Acknowledgements.* This research has made use of the SIMBAD bibliographic facility and the authors wish to acknowledge their gratitude for the same. Thanks are also due to the ADS(Astrophysics Data System) services of NASA which were used frequently during the course of this work. The ISO Spectral Analysis Package (ISAP) is a joint development by the LWS and SWS Instrument Teams and Data Centers. R.S. would like to thank J. S. Nathan for help in the installation of IRAF and IUETOOLS packages. P.G.L. acknowledges support from grant AYA2003–09499, financed by the Spanish Ministerio de Ciencia y Tecnología.

# Online Material

**Table 4.** Ionic concentrations and chemical abundances in Me 2–1. Wavelength in Angstrom for all values of $\lambda$ above 1000, otherwise in $\mu$m.

| Ion | $\lambda$ | I | T | $N_{\rm ion}/N_{\rm p}$ | ICF | $N_{\rm el.}/N_{\rm p}$ |
|---|---|---|---|---|---|---|
| $\rm He^+$    | 5875 | 4.06  | 12 000 | 0.024     |      |          |
| $\rm He^{++}$ | 4686 | 8.77  | 13 500 | 0.076     | 1    | 0.100    |
| $\rm C^+$     | 2325 | 34.   | 11 500 | 2.82(–5)  |      |          |
| $\rm C^{++}$  | 1909 | 821.  | 12 200 | 4.98(–4)  |      |          |
| $\rm C^{+3}$  | 1548 | 1187. | 15 000 | 1.68(–4)  | 1    | 7.0(–4)  |
| $\rm N^+$     | 6584 | 9.9   | 11 500 | 1.35(–6)  |      |          |
| $\rm N^{++}$  | 1750 | 19.7  | 12 600 | 2.40(–5)  |      |          |
| $\rm N^{+3}$  | 1485 | 29.3  | 15 500 | 2.02(–5)  |      |          |
| $\rm N^{+4}$  | 1239 | 25.7  | 18 000 | 5.7(–6)   | 1    | 5.1(–5)  |
| $\rm O^+$     | 3726 | 23.   | 11 500 | 1.17(–5)  |      |          |
| $\rm O^{++}$  | 5007 | 1422. | 13 000 | 2.49(–4)  |      |          |
| $\rm O^{+3}$  | 25.8 | 929.  | 17 000 | 2.48(–4)  | 1    | 5.3(–4)  |
| $\rm Ne^+$    | 12.8 | ≤4.7  | 14 000 | ≤5.9(–6)  |      |          |
| $\rm Ne^{++}$ | 15.5 | 61.4  | 14 000 | 3.83(–5)  |      |          |
| $\rm Ne^{+3}$ | 2424 | 191.  | 17 500 | 3.29(–5)  |      |          |
| $\rm Ne^{+4}$ | 24.3 | 165.  | 18 000 | 1.85(–5)  |      |          |
| $\rm Ne^{+5}$ | 7.65 | 2.96  | 18 500 | 1.12(–7)  | 1    | 9.3(–5)  |
| $\rm S^+$     | 6731 | 2.46  | 11 000 | 1.0(–7)   |      |          |
| $\rm S^{++}$  | 6312 | 1.55  | 12 000 | 1.69(–6)  |      |          |
| $\rm S^{++}$  | 9531 | 31.4  | 12 000 | 1.36(–6)  | 6:   | 9.1(–6): |
| $\rm Ar^{++}$ | 7136 | 9.16  | 12 500 | 5.40(–7)  |      |          |
| $\rm Ar^{+3}$ | 4740 | 5.53  | 14 000 | 6.9(–7)   |      |          |
| $\rm Ar^{+4}$ | 7005 | 1.9   | 17 000 | 1.75(–7)  | 1.2  | 1.6(–6)  |
| $\rm Cl^{++}$ | 5538 | 0.53  | 12 000 | 4.5(–8)   |      |          |
| $\rm Cl^{+3}$ | 8046 | 1.19  | 14 000 | 6.3(–8)   | 1.7  | 1.8(–7)  |
| $\rm K^{+3}$  | 6101 | 0.37  | 13 000 | 1.95(–8)  |      |          |
| $\rm K^{+4}$  | 4163 | 0.15  | 17 000 | 2.0(–8)   |      |          |
| $\rm K^{+5}$  | 6228 | 0.21  | 18 000 | 4.6(–9)   | 1.1: | 5.0(–8): |
| $\rm Mg^{+3}$ | 4.49 | 17.5  | 18 000 | 7.9(–6)   |      |          |
| $\rm Mg^{+4}$ | 5.61 | 15.2  | 18 000 | 3.9(–6)   | 1.7: | 2.0(–5): |
| $\rm Si^{++}$ | 1892 | 2.4   | 11 600 | 3.5(–6)   |      |          |

Intensities (I) given with respect to $H_\beta = 100$.

**Table 7.** The emission line fluxes (H$_\beta$ = 100).

| Label | Line | Model flux | Obsd. flux (dereddened) | Label | Line | Model flux | Obsd. flux (dereddened) |
|---|---|---|---|---|---|---|---|
| TOTL(N 5) | 1240 A | 17.152 | 25.720 | Cl 3 | 5538 A | 0.844 | 0.530 |
| REC(C 2) | 1335 A | 20.298 | 19.340 | Fe 7 | 5722 A | 0.176 | 0.080 |
| TOTL(O 4) | 1402 A | 33.473 | 87.610 | N 2 | 5755 A | 0.222 | 0.220 |
| TOTL(N 4) | 1486 A | 26.800 | 29.260 | He 1 | 5876 A | 3.524 | 4.060 |
| TOTL(C 4) | 1549 A | 1193.302 | 1187.000 | Ca 5 | 6087 A | 0.804 | 0.150 |
| Ne 5 , C 3 | 1575, 77 A | 0.6 , 5.4 | 10.090 | O 1 | 6300 A | 0.037 | 1.620 |
| Ne 4 | 1602 A | 3.159 | 16.510 | S 3 | 6312 A | 1.505 | 1.550 |
| He 2 | 1640 A | 590.967 | 585.850 | O 1 | 6363 A | 0.011 | 0.560 |
| TOTL(O 3) | 1665 A | 42.890 | 45.800 | Ar 5 | 6435 A | 1.129 | 0.990 |
| TOTL(N 3) | 1750 A | 17.816 | 19.700 | N 2 | 6548 A | 3.086 | 3.420 |
| TOTL(Si 3) | 1888 A | 28.517 | 30.660 | H 1 | 6563 A | 282.183 | 321.430 |
| TOTL(C 3) | 1909 A | 839.670 | 821.960 | N 2 | 6584 A | 9.108 | 9.860 |
| C 3 | 2297 A | 5.522 | 14.570 | He 1 | 6678 A | 0.910 | 1.720 |
| TOTL(C 2) | 2326 A | 38.553 | 33.980 | S II | 6716 A | 2.022 | 1.800 |
| Ne 4 | 2424 A | 84.589 | 191.110 | S II | 6731 A | 2.867 | 2.460 |
| He 2 | 2511 A | 11.882 | 12.750 | Ar 5 | 7005 A | 2.421 | 1.900 |
| He 2 | 2733 A | 19.474 | 19.430 | He 1 | 7065 A | 1.159 | 1.300 |
| He 2 | 3203 A | 36.679 | 36.130 | Ar 3 | 7135 A | 10.538 | 9.160 |
| O 3 | 3265 A | 0.948 | 1.830 | He 2 | 7178 A | 1.429 | 0.850 |
| He 1 | 3355 A | 0.019 | 4.460 | Ar 4 | 7237 A | 0.124 | 0.350 |
| Ne 5 | 3426 A | 58.361 | 138.190 | Ar 4 | 7263 A | 0.139 | 0.250 |
| O II | 3726 A | 23.150 | 23.120 | Cl 4 | 7532 A | 0.479 | 0.290 |
| O II | 3729 A | 12.050 | 18.690 | Ar 3 | 7751 A | 2.543 | 2.300 |
| H 1 | 3734 A | 3.870 | 2.950 | Cl 4 | 8047 A | 0.960 | 1.190 |
| H 1 | 3750 A | 4.804 | 3.200 | H 1 | 8598 A | 1.056 | 0.530 |
| H 1 | 3771 A | 6.035 | 4.330 | H 1 | 8665 A | 1.310 | 0.940 |
| H 1 | 3798 A | 7.758 | 4.770 | H 1 | 8750 A | 1.624 | 0.940 |
| He 1 | 3820 A | 0.290 | 0.490 | H 1 | 8863 A | 2.040 | 1.240 |
| H 1 | 3835 A | 8.037 | 7.010 | H 1 | 9015 A | 2.619 | 1.360 |
| S II | 4070 A | 0.775 | 1.590 | S 3 | 9069 A | 13.549 | 12.170 |
| S II | 4078 A | 0.250 | 0.420 | H 1 | 9229 A | 2.704 | 2.420 |
| C 3 | 4159 A | 0.015 | 0.260 | He 2 | 9345 A | 2.981 | 2.240 |
| C 2 | 4267 A | 0.336 | 0.410 | H 1 | 1.005 m | 5.510 | 5.420 |
| H 1 | 4340 A | 47.440 | 48.670 | He 2 | 1.012 m | 22.765 | 20.550 |
| TOTL(O 3) | 4363 A | 22.280 | 22.080 | Si 7 | 2.483 m | 0.000 | <2.650 |
| He 1 | 4388 A | 0.143 | 0.190 | H 1 | 4.051 m | 7.304 | 4.310 |
| He 1 | 4471 A | 1.180 | 1.240 | Ca 7 | 4.086 m | 1.543 | <12.930 |
| He 2 | 4686 A | 85.612 | 87.730 | Mg 4 | 4.485 m | 22.523 | 17.540 |
| Ar 4 | 4711 A | 5.778 | 7.900 | Ar 6 | 4.528 m | 11.800 | <10.130 |
| Ne 4 | 4720 A | 0.728 | 1.750 | Mg 7 | 5.502 m | 0.000 | <18.170 |
| Ar 4 | 4740 A | 5.409 | 5.530 | Mg 5 | 5.608 m | 12.233 | 15.190 |
| O 3 | 4959 A | 526.738 | 485.760 | Si 7 | 6.491 m | 0.000 | <10.640 |
| O 3 | 5007 A | 1585.481 | 1422.210 | Ne 6 | 7.650 m | 3.927 | 2.960 |
| Fe 6 | 5177 A | 0.154 | 0.130 | Ne 2 | 12.81 m | 0.296 | <4.740 |
| Fe 3 | 5270 A | 0.011 | 0.040 | Mg 5 | 13.52 m | 1.006 | <3.150 |
| N 4 | 5209 A | 0.000 | 0.130 | Ne 5 | 14.32 m | 171.256 | 122.490 |
| Cl 4 | 5324 A | 0.030 | 0.060 | Ne 3 | 15.55 m | 67.821 | 61.350 |
| He 2 | 5412 A | 6.894 | 6.360 | Ne 5 | 24.31 m | 136.497 | 165.360 |
| Cl 3 | 5518 A | 0.874 | 0.570 | O 4 | 25.88 m | 813.774 | 929.370 |

Absolute H$_\beta$ flux model: $8.32 \times 10^{-12}$ erg cm$^{-2}$ s$^{-1}$ Obsn: $8.93 \times 10^{-12}$ erg cm$^{-2}$ s$^{-1}$.

**Notes**: "A" in Col. "Line" signifies Angstrom; "m" signifies $\mu$m. In Col. "Label", we have followed the notation used by Cloudy for atoms and ions; this will make identifying a line in Cloudy's huge line list easy. Neutral state is indicated by "1" and singly ionized state by "2" etc., "TOTL" typically means the sum of all the lines in the doublet/multiplet; or it could mean sum of all processes: recombination, collisional excitation, and charge transfer. "REC" indicates radiative recombination line; for such cases, we have added the ionic name within parentheses for convenience though Cloudy does not. Some elements are represented by usual notation as per Cloudy. A symbol "<" in the Col. "Obsd. flux" indicates that these lines have not been detected by ISO and the values given are upper limits.